\documentclass[twocolumn,10pt]{IEEEtran}
\usepackage{bm,cite,float,amsmath,amssymb,amsthm}

\usepackage{algorithm}
\usepackage[noend]{algpseudocode}
\usepackage{indentfirst}
\usepackage{xcolor}
\usepackage{graphicx}
\usepackage{subfigure} 
\usepackage{booktabs}    

\usepackage{lipsum}
\usepackage{mathtools}
\usepackage{cuted}

\makeatletter
\def\BState{\State\hskip-\ALG@thistlm}
\makeatother

\usepackage{amssymb}
\usepackage{amsmath}
\usepackage{graphicx}
\usepackage{cite}

\usepackage{balance}
\usepackage[utf8]{inputenc}

\IEEEoverridecommandlockouts

\usepackage{hyperref}
\usepackage{graphicx,epstopdf}
\usepackage{epsfig}	
\usepackage{amsfonts,balance}
\usepackage{bbm}
\floatname{algorithm}{Algorithm}

\usepackage{lipsum}

\newtheorem{theorem}{Theorem}

\newtheorem{corollary}{Corollary}

\makeatletter
\def\ScaleIfNeeded{%
\ifdim\Gin@nat@width>\linewidth \linewidth \else \Gin@nat@width
\fi } \makeatother

\definecolor{backgroundcolor}{RGB}{255, 255, 255}
\pagecolor{backgroundcolor}

\title{Task-Oriented Metaverse Design in the 6G Era}
\author{Zhen~Meng,~\IEEEmembership{Student Member,~IEEE,}
	Changyang~She,~\IEEEmembership{Senior Member,~IEEE,}\\
	Guodong~Zhao,~\IEEEmembership{Senior Member,~IEEE,}
	Muhammad~A.~Imran,~\IEEEmembership{Fellow,~IEEE}, Mischa~Dohler,~\IEEEmembership{Fellow,~IEEE}, Yonghui Li,~\IEEEmembership{Fellow,~IEEE}, and Branka Vucetic,~\IEEEmembership{Life Fellow,~IEEE}

}
\begin{document}
\maketitle
\maketitle

\begin{abstract}
As an emerging concept, the Metaverse has the potential to revolutionize the social interaction in the post-pandemic era by establishing a digital world for online education, remote healthcare, immersive business, intelligent transportation, and advanced manufacturing. The goal is ambitious, yet the methodologies and technologies to achieve the full vision of the Metaverse remain unclear.  In this paper, we first introduce the three infrastructure pillars that lay the foundation of the Metaverse, i.e., human-computer interfaces, sensing and communication systems, and network architectures. Then, we depict the roadmap towards the Metaverse that consists of four stages with different applications. To support diverse applications in the Metaverse, we put forward a novel design methodology: task-oriented design, and further review the challenges and the potential solutions. In the case study, we develop a prototype to illustrate how to synchronize a real-world device and its digital model in the Metaverse by task-oriented design, where a deep reinforcement learning algorithm is adopted to minimize the required communication throughput by optimizing the sampling and prediction systems subject to a synchronization error constraint.


\end{abstract}



%
\maketitle

\section{Introduction}
The Metaverse is a digital world that will revolutionize the interactions among humans, machines, and environments by providing a shared, unified, perpetual, and inter-operable realm for participants from all over the world~\cite{lee2021all}. The digital world could be a pure virtual space or a digital mirror of the physical world that has the ability to reprogram the physical world in real time. It lays the foundation for the evolution of different vertical industries including education, entertainment, healthcare, manufacturing, transportation, and immersive business. This ambitious vision brings significant challenges to the development of next-generation communication networks. It is natural to raise the following questions:

{\bf{Q1: Is the available infrastructure sufficient for the Metaverse?}}

To support an application in the Metaverse, the system needs to execute a sequence of interdependent tasks. A task is an activity that needs to be completed within a period of time or by a deadline, such as the pose and eye tracking, positioning, haptic control and feedback, and semantic segmentation. State-of-the-art infrastructure cannot meet the requirements of diverse emerging applications and tasks in the Metaverse. Specifically, existing input/output systems, like the touch screen, keyboards, and mouses, are inconvenient in supporting new tasks. Thus, new Human-Computer Interface (HCI), including Virtual/Augmented Reality (VR/AR), Tactile Internet, and brain-computer interface, will lay the foundation for the Metaverse. Sensing and communication technologies play critical roles in providing timely feedback and seamless connections in the Metaverse with a real-world counterpart. To reduce the infrastructure cost, one promising approach is to exploit widely deployed mobile networks for both sensing and communications. Furthermore, the Sixth Generation (6G) networks will bridge new HCI and sensing \& communication systems. Due to the long propagation delay, executing all tasks on a global server cannot meet the latency requirements of tasks.  A new multi-tier network architecture that can coordinate computing, communication, and storage resources at the end-user devices, edge/local servers, and global servers efficiently is essential for supporting interdependent tasks in the Metaverse~\cite{multi-tier}. In summary, HCI, sensing and communication technologies, as well as network architectures will serve as the three pillars of the Metaverse. 

Even with the above infrastructure, supporting emerging applications in the Metaverse is not straightforward.

{\bf{Q2: How to guarantee the Key Performance Indicators (KPIs) of diverse applications/tasks in the Metaverse?}}

The highly integrated and multifaceted demands of applications in the Metaverse impose stringent requirements on KPIs that are much more diverse than the KPIs defined in the three typical scenarios in the Fifth Generation (5G) mobile communication standard, i.e., Enhanced Mobile Broadband (eMBB), Ultra-Reliable Low-Latency Communications (uRLLC), and Massive Machine Type Communications (mMTC)~\cite{3GPP}. Further considering that one application consists of multiple tasks, to meet the specific requirements of the application in the Metaverse, we should analyze the KPIs at task level, referred to as task-oriented KPIs. For example, to generate haptic feedback, the system should meet the Just Noticeable Difference (JND) constraint, which is the minimum difference between two force signals that is noticeable to users. The network functions and communication KPIs in 5G networks are task agnostic and hence cannot guarantee task level KPIs.

The existing communication network design approach divides the whole system into multiple sub-modules for separate optimization and cannot break the barriers among the sub-modules. As a result, it is difficult to provide End-to-End (E2E) performance guarantees. To support the Metaverse in 6G mobile networks, we should revisit the following questions: 

{\bf{Q3: What are the issues with the existing design approaches? Do we need new design methodologies in 6G?}}

To improve E2E performance, cross-system design has been investigated in the existing literature~\cite{9493202,liu2022integrated,edgemeta}. To guarantee the control performance with stochastic wireless channels and limited communication resources in mission-critical applications, a predictive control and communication co-design system was introduced in~\cite{9493202}, where scheduling policy and transmission power are jointly optimized. To achieve substantial gains in spectral, energy, hardware and cost efficiency with mMTC, Integrated Sensing and Communication (ISAC) was developed in \cite{liu2022integrated} to support sensing and communications simultaneously. Considering that end-user devices have limited computing, communication, and storage resources, a cloud-edge-end computing framework-driven solution was introduced in~\cite{edgemeta}. 
However, the cross-system design problems are in general non-convex or NP-hard, and novel design methodologies for real-time implementation are in urgent need.

In this work, we introduce the task-oriented design for the Metaverse in the 6G era. The major contributions of this paper are summarized as follows: 1) We holistically illustrate the three infrastructure pillars that the Metaverse will be built upon, and depict the roadmap toward the full vision of the Metaverse. 2) We comprehensively review the challenges of task-oriented design in the Metaverse; To tackle these challenges, we put forward potential solutions from a system-level perspective. 3) We build a prototype to demonstrate the task-oriented design. The goal of the task is to synchronize a real-world device and its digital model in the Metaverse. 






\section{Pillars of the Metaverse}
\begin{figure*}
\centering
\includegraphics[scale=0.42]{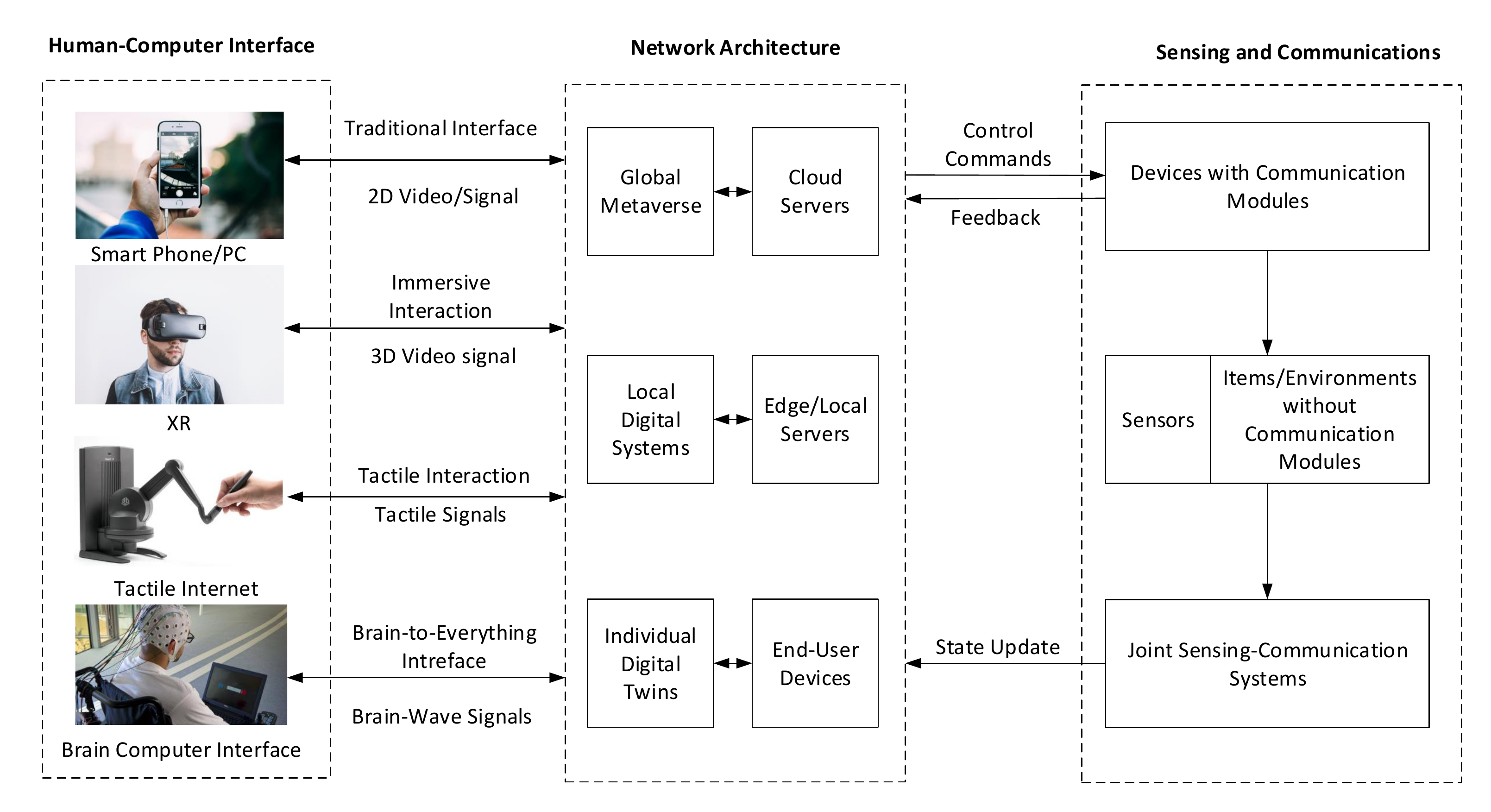}
\caption{Three pillars of the Metaverse, i.e., human-computer interface, sensing and communications and network architecture.}
\label{pillars}
\end{figure*}
In this section, we review the Metaverse infrastructure and its connections to the task-oriented design. As shown in Fig.~\ref{pillars}, it consists of three pillars: the human-computer interface (HCI),  sensing \& communications, and network architecture. 

\subsection{Human-Computer Interface}
Different from existing input/output systems that are designed to process video and audio signals, future HCI should be carefully designed to support new tasks in the Metaverse.


\subsubsection{XR Head-Mounted Devices} The development of XR devices has greatly improved the user experience by identifying the mobility of the head-mounted device and rendering the three-dimensional (3D) video accordingly. Existing XR systems mainly focus on downlink video streaming. To further enable eye contact and expression reconstruction in the Metaverse, eye tracking and 3D modeling techniques should be integrated into XR systems. By predicting the moving direction of eyes, the XR system can render and transmit the field-of-view to be requested by users. Thus, we can improve the trade-off between data rate and latency in wireless XR.


\subsubsection{Tactile Devices} Tactile devices are essential for supporting haptic feedback in the Metaverse. With a large number of tactile sensors and actuators, it is possible to recognize users' poses and gestures. Once the user hits a virtual item in the Metaverse, the tactile devices generate feedback to users via vibrations and resistance. Most existing tactile devices cannot provide tactile feedback for the entire human body. Several issues remain open in the development of whole-body tactile devices: 1) battery-life time of wearable devices is limited; 2) low-complexity graph signal processing that takes the topology of the sensors/actuators is not available; 3) the actuators should be controlled by engines and algorithms to mimic the tactile experience, which remains an open challenge.


\subsubsection{Brain-Computer Interface} The brain-computer interface can be used for emotion recognition and reconstruction in the Metaverse. Existing brain-computer interfaces suffer from low classification accuracy and long processing delay. Due to these issues, the brain-computer interface may not be able to work as the stand-alone human-computer interface in the near future, but it may assist VR devices or tactile devices to improve the users' experience, as demonstrated in early trials by Meta.

\subsubsection{Combination of Different Human-Computer Interfaces} Different human-computer interfaces have different data structures, generate responses in different time scales, and may support different tasks in one application. Developing a system that manages multiple human-computer interfaces brings unprecedented challenges, and is crucial for improving the users' experience in the Metaverse. To enable the interactions among users with different types of devices, new standards are needed.


\subsection{Sensing and Communications}
Sensing and communication technologies enable timely state updates of real-world devices/environments in the Metaverse. Thus, they are critical to the establishment of the digital world. 



\subsubsection{Devices with Communication Modules} Smart devices equipped with communication modules can update their states to the Metaverse. For example, a real-world robotic arm measures the angles, speeds, forces, and torques of the joints and sends the states to a server for reconstructing the digital robotic arm. As the number of devices increases, the communication resources become the bottleneck of the Metaverse. Improving the trade-off between the communication resource utilization efficiency and the synchronization accuracy/information freshness is a challenging problem. 



\subsubsection{Environments without Communication Modules} Some entities in real-world environments do not have communication modules, such as trees, buildings, pedestrians, etc. To collect their states in the digitally twinned Metaverse, we need a large number of external sensors or cameras. For example, Instant-Nerf is a neural rendering model developed by NVIDIA that can render 2D photos into 3D scenes in a few milliseconds~\cite{mueller2022instant}. To further understand the environment, semantic segmentation is crucial~\cite{long2015fully}. Nevertheless, most of the existing segmentation algorithms require a considerable amount of computation resources, and the processing time remains the bottleneck of real-time interactions.

\subsubsection{Joint Sensing-Communication Systems} The cost of deploying and operating a large number of sensors and cameras could be extremely high. By integrating communication and sensing functionalities into widely deployed mobile networks, it is possible to reduce the cost. Thus, the Integrated Sensing and Communication (ISAC) system is a practical approach that collects real-world information for the Metaverse~\cite{liu2022integrated}. Note that there are tradeoffs between KPIs of different tasks and the resource utilization efficiency of ISAC systems, but a universal design framework for different tasks is still missing.

























\subsection{Network Architecture}
6G networks will bridge HCI and sensing \& communication devices, and the Metaverse. \subsubsection{Multi-tier Computing Architecture} Developing the Metaverse on a global server for all the users and devices around the world would be very challenging due to long communication delay and the limited communication throughput. Multi-tier computing is believed to be a promising architecture that can coordinate interdependent tasks in the Metaverse by exploiting distributed computing, storage, and communication resources in central servers, local servers, and end-user devices \cite{multi-tier}. Meanwhile, the multi-tier computing  raises new challenges in 6G core networks and radio access networks (RANs).

\subsubsection{Core Networks} 5G core networks manage resources and quality-of-service at the application level. The session management function will create a new protocol data unit session when there is a new service request. How to coordinate multiple tasks for one application remains unclear. To address this issue, several promising techniques have been investigated in the existing literate: 1) The authors of \cite{qin2021semantic} developed a semantic-effectiveness plane task-level information processing. 2) In \cite{chaccour2022data}, the authors built a knowledge pool for reasoning-driven AI-native systems that enable online learning and fast inference of different network functions.

\subsubsection{Radio Access Networks} Most new HCI and sensing \& communication devices will access to  RANs for better user experience and flexible deployment. As a result, 6G RANs should support massive devices with diverse KPI requirements. To improve user experience in real-time interactions, ISAC is a promising technology that exploits a shared multi-antenna system and advanced signal processing algorithms for data transmission and environment sensing~\cite{liu2022integrated}. In addition, a space-air-ground-sea integrated network is promising for enabling seamless connectivity for global interactions in the Metaverse~\cite{edgemeta}. As the tasks and applications in the Metaverse evolve over time, Open-Radio Access Networks (O-RAN) with programmable network functions can reduce the cost for network deployment and upgrades significantly~\cite{chaccour2022data}.

\section{Road Map Towards The Metaverse}
\begin{figure*}
\centering
\includegraphics[scale=0.50]{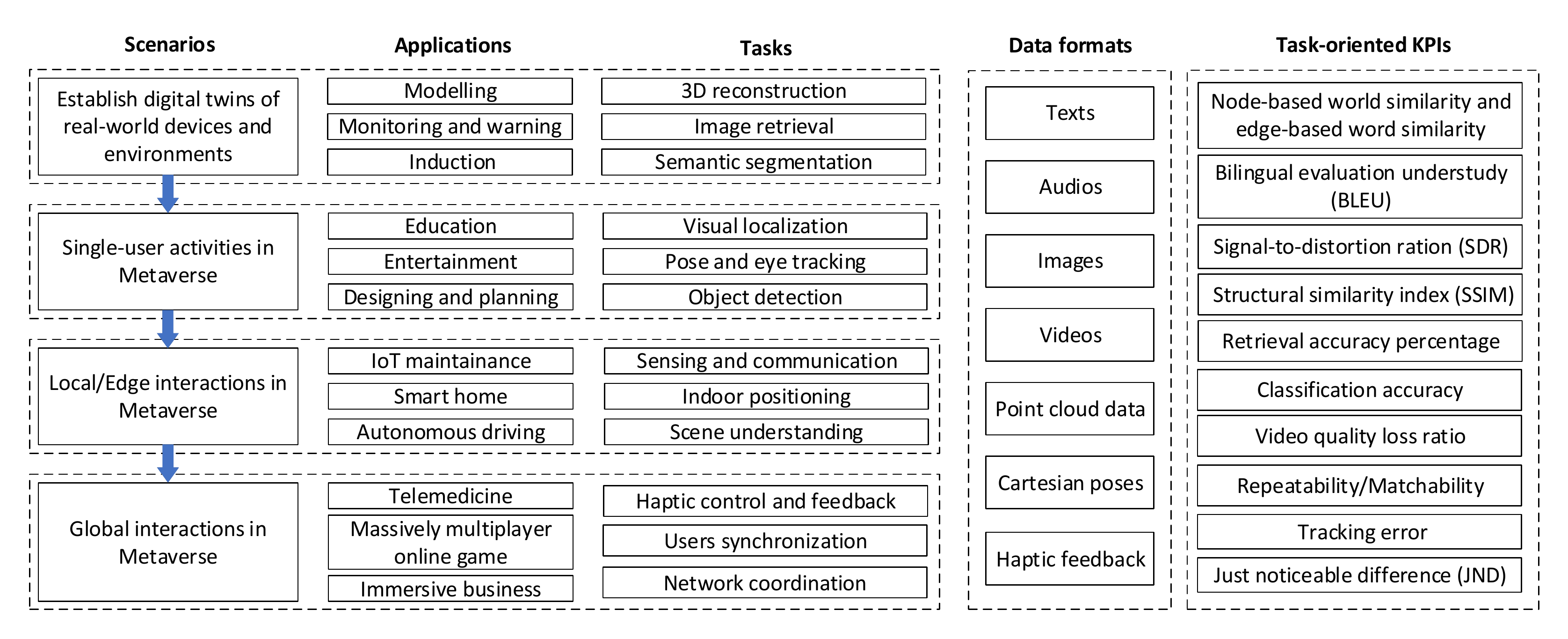}
\caption{Road map towards the full-vision of the Metaverse.}
\label{frame}
\end{figure*}

In this section, we discuss the road map towards the full vision of the Metaverse as illustrated in Fig. \ref{frame}.
\subsection{Establish Multi-tier Metaverse in Multi-tier Architecture}
The first step toward the Metaverse is to build digital worlds. There are three types of digital worlds: 1) an imaginary environment that does not have a real-world counterpart; 2) a digital twin of a real-world environment; and 3) a digital world overlays on the physical world or even has the ability to reprogram the physical world in real time. To provide quick responses from the Metaverse to users, we need to use the computing, storage, and communication resources of a local server or the end-user device. Then, the states of the user or its digital model are updated to the global sever for synchronization. The multi-tier network architecture lays the foundation for building multi-tier digital worlds that support real-time interactions among users from all over the world.

\subsection{Single-User Activities in the Metaverse} 
For single-user activities, all the virtual objects and environments can be built on the end-user device, e.g., a personal computer. With the help of a variety of HCI, the user can interact with everything in the digital world, where several applications in education, entertainment, designing, and planning become possible. For example, by synchronizing users' actions with their digital models in the Metaverse, the users can operate virtual objects and create virtual content (e.g., driving a vehicle or painting). Nevertheless, establishing a digital world on the end-user device is not easy, as the device has limited computing and storage resources. Thus, the processing delay could be the bottleneck for real-time interactions. To address this issue, low-complexity 3D reconstruction and segmentation algorithms are in urgent need.



\subsection{Local Interactions in the Metaverse} 
In some private networks or local area networks, information is exchanged among devices and users in a small area. In these scenarios, all the devices and human users can interact with each other via a local Metaverse. For example, in a smart factory, sensors monitor manufacturing processes and update their states to the local sever, where a digital twin of the factory is built~\cite{_2022_industry}. In the digital factory, it is possible to simulate the outcomes of different actions. If an accident is detected in the simulation, the local server sends commands to actuators to stop the processes in an anticipatory manner. In addition, the data is stored and processed in the local server that is not connected to the Internet. Therefore, this approach can protect users' privacy and avoid security issues.



\subsection{Global Interactions in the Metaverse} 
The ultimate goal of the Metaverse is to support global interactions for a large number of users using different types of HCI. In this stage, remote healthcare, immersive business, and online education will be possible. Latency remains  one of the major issues for global interactions. Specifically, the propagation delay is inevitable in long-distance communications. Stochastic network congestion leads to long queueing delay. In addition, the re-transmission scheme in the existing Transmission Control Protocol/Internet Protocol brings significant latency. Although some interesting ideas have been put forward to achieve real-time interactions \cite{meng2022sampling}, the implementation in large-scale networks remains a challenging goal.

\section{Task-Oriented Design framework} 

In this section, we propose a task-oriented design framework. In general, there are three types of tasks in the Metaverse: 1) environment sensing or measurements with HCI for constructing the digital world, 2) data/signal processing for understanding, prediction, inference, and generating feedback, and 3) communications for information exchange among human users, machine-type devices, and servers. Let us take a virtual conference as an example to illustrate the tasks. As body language plays an important role in virtual conference, the HCI executes the task of identifying the human pose and eye tracking. After that, the computer/sever processes the measured/estimated data. In this stage, typical tasks include segmentation, object detection, and rendering of 3D scenes. Finally, based on the interaction between the avatars in the Metaverse (such as shaking hands), the feedback is generated and sent back to users by the Tactile Internet.

\subsection{Challenges of Task-Oriented Design}
\subsubsection{Data Structures}
The data structure of a task depends on the HCI or environment sensing technologies. Traditional speech and image signals are represented by time-series data and the red-green-blue (RGB) model, respectively.  Nevertheless, spatial correlation is critical for the tactile signals and brainwave signals, and relies on the topology of the sensors. The topology information is useful in signal processing, and may facilitate the execution of tasks. For example, the signals generated by a radar system or depth-sensing camera, such as point-cloud data are converted into 3D tensors in the Euclidean space before they can be processed by convolutional neural networks. This procedure causes additional computational overhead and processing delay. To reduce overhead and improve the performance of a task, the authors of \cite{qi2017pointnet} developed a PointNet to handle a range of tasks in environment sensing, such as 3D shape classification and segmentation. Nevertheless, a widely accepted standard for data storage, processing, or communications is still missing in the Metaverse, and it will lay the foundation for immersive interactions among human users, machine-type devices, and environments. 


\subsubsection{Task-Oriented KPIs}
Diverse tasks in the Metaverse have stringent requirements on a range of KPIs, which are still difficult to fulfill. For example, in applications that require low-latency feedback, the user-experienced delay should be close to zero, but the propagation delay could be up to dozens of milliseconds when the communication distance is hundreds or thousands kilometers. Furthermore, the KPIs defined in the 5G standard, such as throughput, latency, and reliability, are not the same as the task-oriented KPIs as illustrated in Fig. \ref{frame}. For instance, in haptic communications, it is natural to raise a question: Do we really need to guarantee the $99.999$\% reliability in communication systems in order to achieve the target JND? The impact of network resource allocation on the task-oriented KPIs remains unclear, and there is no theoretical model or closed-form expression that can quantify their relationships. To overcome this difficulty, we need novel design methodologies. 

\subsubsection{Multi-Task Processing and Coordination}
With the multi-tier network architecture, tasks of an application may be executed by the end-user device, an edge/local server, or the cloud server. The offloading and coordination of multiple tasks is not trivial since they are interdependent. For some highly interactive applications, the end-user device senses the behavior of the user, and then communicates with the local server, where the feedback is generated. Finally, the local servers synchronize the states of users in a cloud server. Delays or packet losses in any of the tasks will have a serious impact on the overall performance of the application. To provide satisfactory user experience in the Metaverse, we need to break the barriers among sensing, communication and computing systems, and jointly design the whole network.

\subsection{Potential Solutions}

\subsubsection{Cross-System Design} 
Existing HCI, sensing, communication, and computing systems are developed separately. This design approach leads to sub-optimal solutions, brings extra communication overhead for coordinating multiple tasks, and can hardly meet the task-oreinted KPIs. To address these issues, a cross-system design has been investigated in the existing literature. There are several existing cross-system design approaches. (1) As shown in~\cite{qin2021semantic}, when dealing with reconstruction tasks including in-text sentences, sounds, images, and point cloud data, by joint source and channel coding, it is possible to achieve a better quality of service at low signal-to-noise ratios. Nevertheless, complicated coding schemes may bring extra processing delay, which remains an issue in ultra-low latency communications. (2) Considering the cost of deploying a large number of sensors, integrating sensing into communication systems is a promising approach, as cellular networks have been widely deployed \cite{liu2022integrated}. By utilizing communication signals in environmental sensing, cellular networks can support a variety of tasks, such as localization, object detection, and health monitoring. (3) Given the fact that state observations are outdated in some tasks, the Metaverse needs to respond to users' actions in an anticipatory manner. To achieve this goal, prediction and communication co-design is promising, especially for applications in the Tactile Internet that requires ultra-low latency~\cite{meng2022sampling}.

It is worth noting that cross-system problems are in general very complicated and may not have well-established models. As a result, most of the existing analytical tools and optimization algorithms are not applicable.

\subsubsection{Domain-Knowledge-Assisted Deep Learning}
To solve the above cross-system design problems, data-driven deep learning methods are promising, as they do not rely on theoretical models or assumptions. However, straightforward applications of deep learning may not generalize well with diverse task-oriented KPIs and data structures \cite{she2020tutorial}. To address this issue, one should exploit domain knowledge in feature engineering, sample selection, value function design, etc. When deep learning is adopted in task-oriented design, there are three major issues. (1) Most of the existing deep neural networks work well in small-scale problems. As the scale of the problem increases, the training/inference time increases rapidly. In the Metaverse, there could be millions or billions of users and devices, and thus scalability remains an open issue. (2) To support various tasks in different sensing and communication environments, deep learning algorithms trained on a data set should achieve good performance in different use cases after a few steps of fine-tuning. This generalization ability is critical for using deep learning in the Metaverse. (3) Most deep learning algorithms do not offer a performance guarantee in terms of classification or regression accuracy. But the KPIs required by some mission-critical tasks are sensitive to the outcomes of learning algorithms. Improving the safety of deep/reinforcement learning algorithms by exploiting domain knowledge is a promising and vital approach.


\subsubsection{Universal Design}
The Metaverse aims to provide better interactions among users with different cultural backgrounds and health conditions (e.g., careers, nationalities, abilities or disabilities, etc.). Different users may have different preferences, habits, and cognition. Meanwhile, they may use different types of HCI devices with different data structures. The diversity of users brings significant challenges in the design of the Metaverse, and the universal design is essential for the success of the Metaverse by considering the diverse needs and abilities of all the users throughout the design process, standardization, and government regulation. For example, a universal design platform named Omniverse can meet user demands from different backgrounds (e.g., artists, developers, and enterprises), where the Universal Scene Description is promising to be the open and extensible standard language for the 3D Internet to eliminate the barriers among different user communities \cite{omi}. Nevertheless, a lot of effort is still needed in the universal design.

\section{A Case Study}
  \begin{figure}
            \centering
            \includegraphics[scale=0.35]{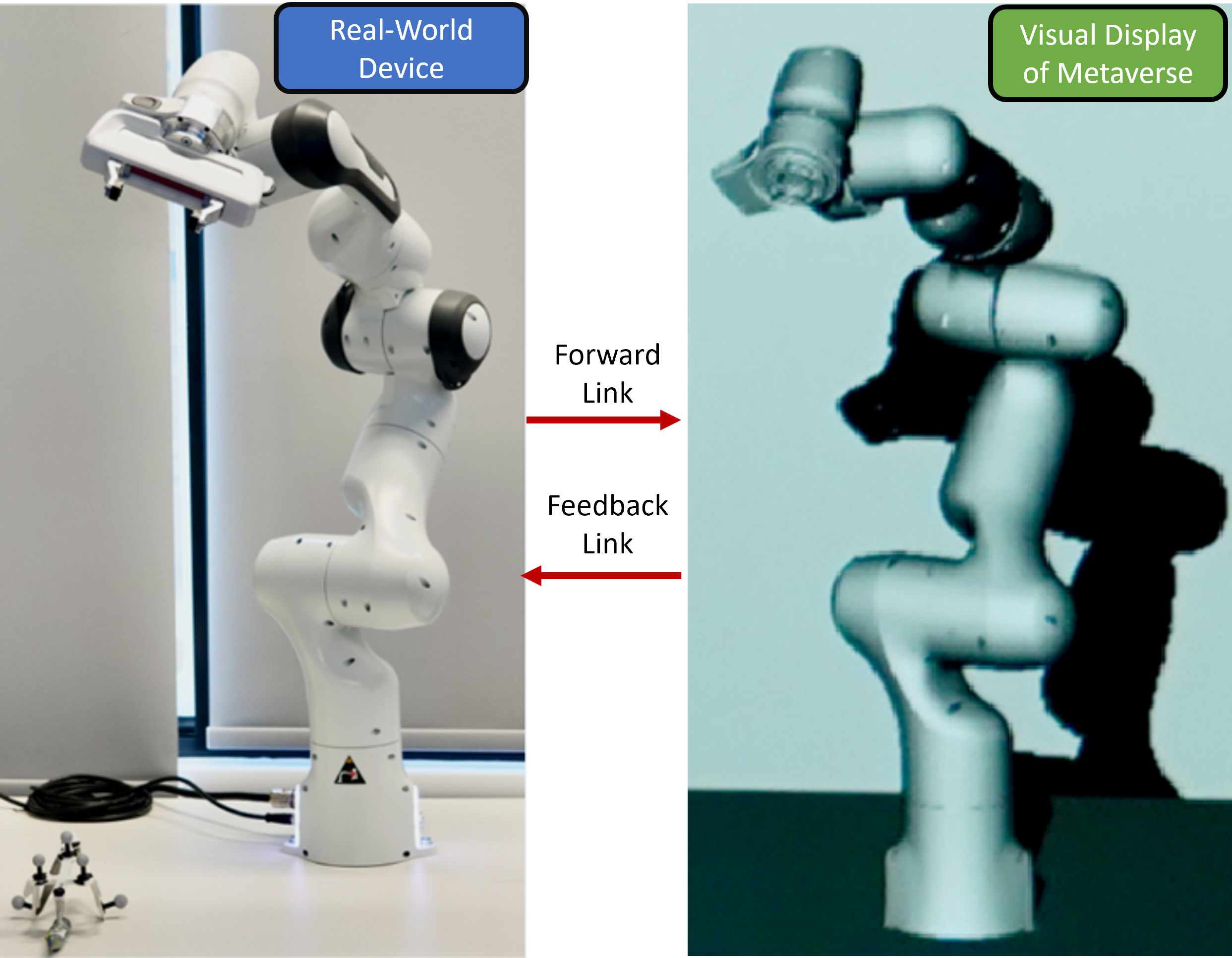}
           \caption{Prototype of synchronizing a real-world robotic arm and its digital model in the Metaverse. The demonstration video is available at \url{https://youtu.be/LCqSGtkrgug}.}
           \label{Prototype}
  \end{figure}

\begin{figure}
            \centering
            \includegraphics[scale=0.5]{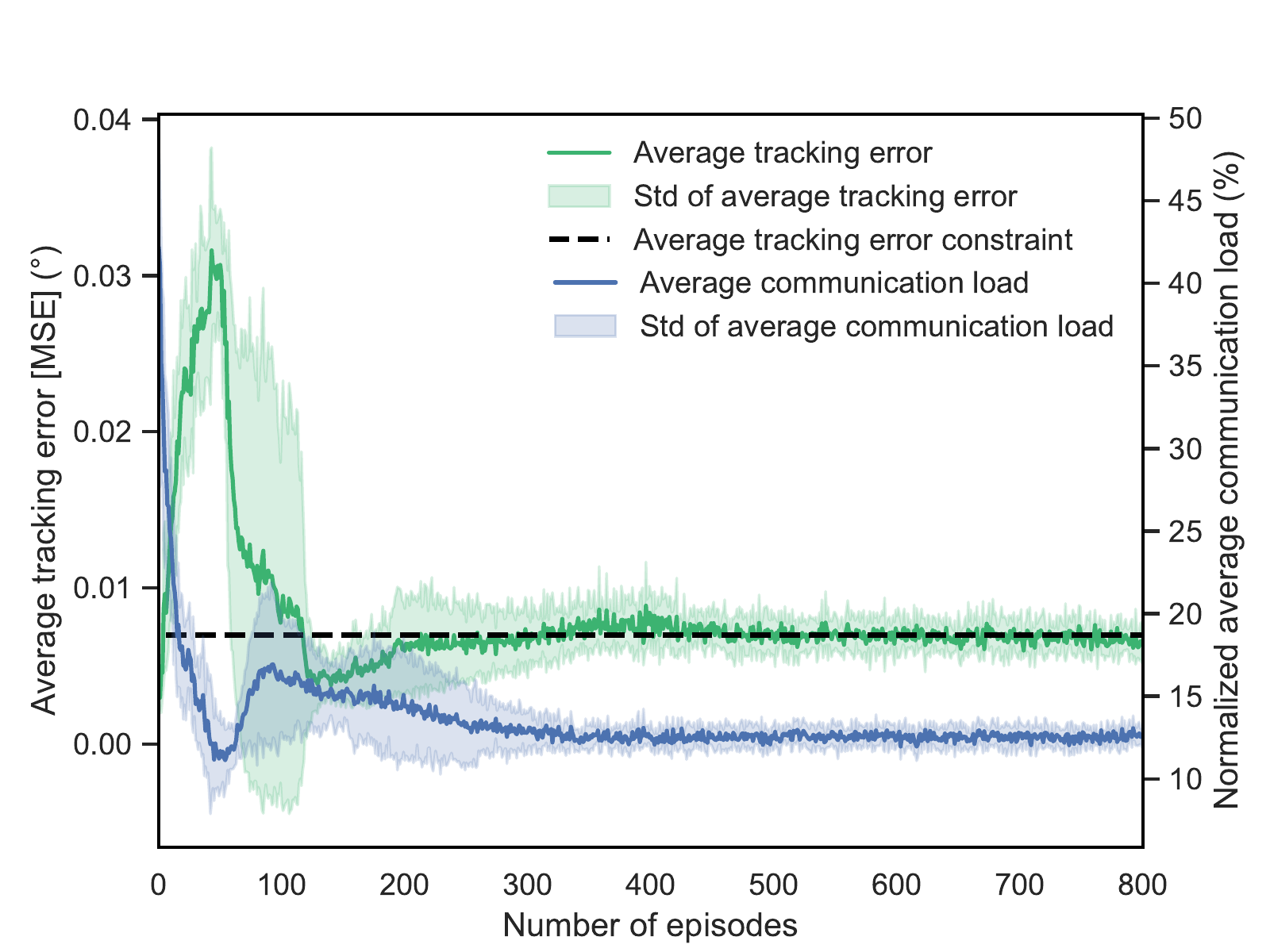}
           \caption{Normalized average communication load and average tracking error in each episode. }
           \label{Fig_loss}
  \end{figure}
\begin{figure}
            \centering
            \includegraphics[scale=0.49]{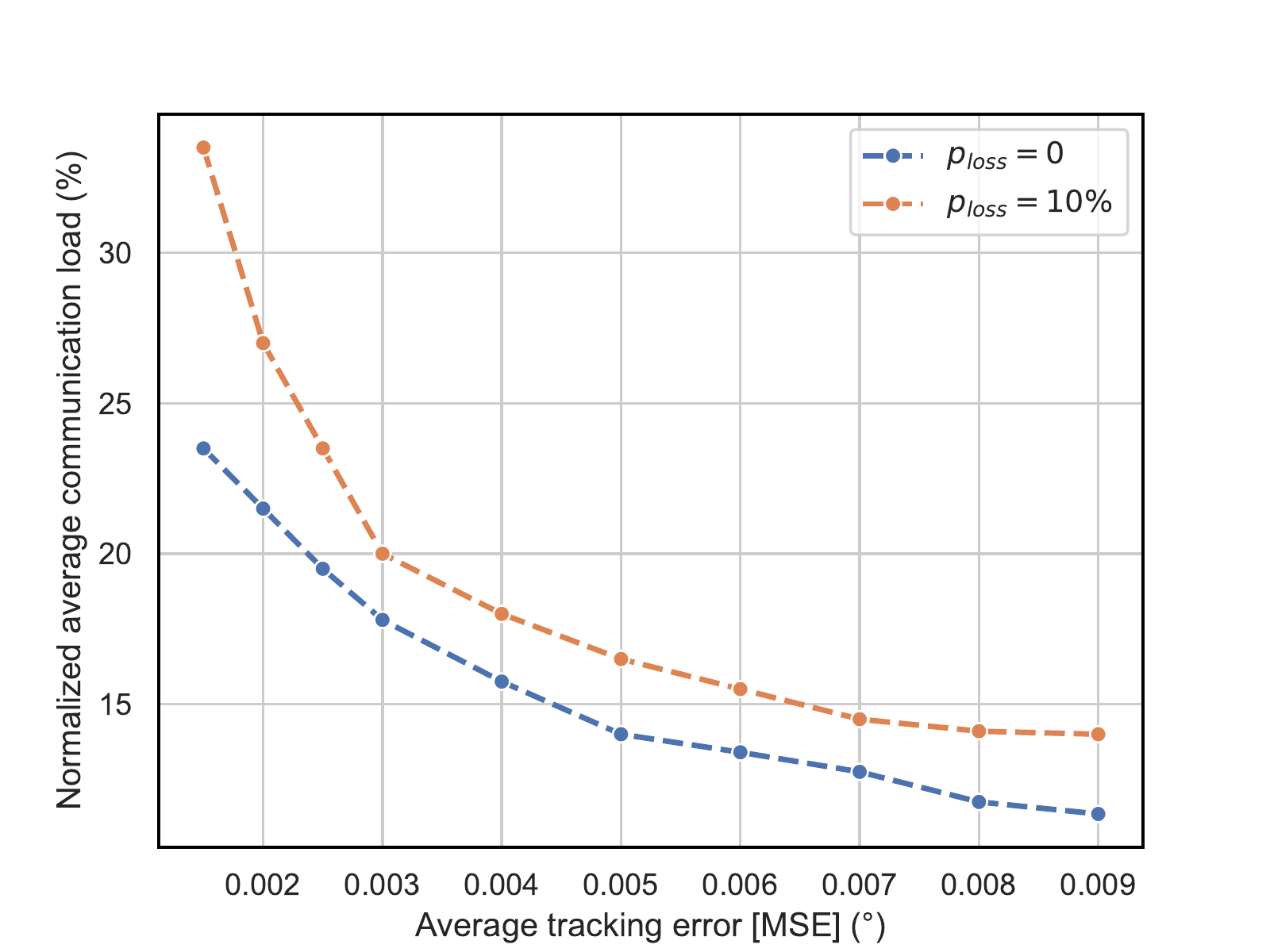}
           \caption{Trade-off between normalized average communication load and average tracking error with different packet loss probabilities.}
           \label{Tradeoff}
  \end{figure}

Timely and accurate synchronization between the real-world device and its digital model is the foundation of the Metaverse. In this section, we show how to jointly optimize the sampling, communication, and prediction modules for the synchronization task shown in Fig.~\ref{frame}. We use domain-knowledge to design a deep reinforcement learning (DRL) algorithm to minimize the communication load subject to an average tracking error constraint. The state is defined as the mean square error (MSE) between the trajectories of the real-world robotic arm and its digital model in the Metaverse. The action includes the prediction horizon and sampling rate. The reward is the communication load. Unlike the latency and reliability constraints in communication system design, our task-oriented design approach aims to guarantee a task-oriented KPI, i.e., the average tracking error. More details about the experiment and the DRL algorithm can be found in \cite{meng2022sampling}.

\subsection{Prototype Setup}
To validate the algorithm, we built a prototype as shown in Fig. \ref{Prototype}, where a virtual robotic arm is synchronized with a physical robotic arm in the real world. Specifically, the sensor attached to the robotic arm measures its trajectory (i.e., angle of the first joint) at the frequency of 1kHz. Then, the measured trajectory is sampled, i.e. decimated, and transmitted to the Metaverse, where the server predicts and reconstructs future trajectory to reduce the latency experienced by the user. Then, the digital model of the robotic arm follows the predicted trajectory and feeds back the prediction results to the real-world robotic arm. Finally, the real robotic arm computes the mean square error (MSE) between the measured trajectory and the predicted trajectory, and a deep reinforcement learning algorithm is applied to adjust the sampling rate and the prediction horizon. For data collection, we consider the motion-controlled robotic arm application, where the real robotic arm is controlled by a human operator. Other details of the system setup and the deep reinforcement learning can be found in \cite{meng2022sampling}.

\subsection{Performance Evaluation}
There are two performance metrics: the average communication load and the average tracking error. In the case without sampling, the packet rate in the communication system is $1,000$~packets/s, which is used to normalize the communication load. For example, if the average packet rate is $150$~packets/s, then the normalized average communication load is $15$~\%. The average tracking error is measured by the average MES between the real-world trajectory and the reconstructed trajectory in the Metaverse.

The results in Fig. \ref{Fig_loss} show the average tracking error and the normalized average communication load in the training stage of the deep reinforcement learning algorithm, where the average tracking error constraint is $0.007^{\circ}$. The results show that the task-oriented design approach can meet the average track error constraint and can reduce the average communication load to $13$\% of the communication load in the system without sampling.


In Fig.~\ref{Tradeoff}, we test the trade-off between the normalized average communication load and the average tracking error, where different packet loss probabilities in the communication system are considered, i.e.,  $p_\text{{loss}} = 0, \ \text{and} \ 10\%$. The results show that with a smaller packet loss probability, it is possible to achieve a better trade-off between the normalized average communication load and the average tracking error. In a communication system with a packet loss probability of $10$~\%, our task-oriented design approach can reduce the normalized average communication load to $27$~\% when the average tracking error constraint is $0.002^{\circ}$. This observation indicates that by adjusting the sampling rate (i.e., the communication load in Fig.~\ref{Tradeoff}), it is possible to meet the requirement of a task in communication systems with high packet loss probabilities, e.g., $10$~\%.

\section{Conclusion and Future Directions}
In this paper, we introduced the three infrastructure pillars and depicted the road map toward the full vision of the Metaverse. Then, we proposed a task-oriented design approach followed by a prototype in a case study. In future 6G standards, we need new network functions for task-level resource management. As the tasks may evolve according to the road map of the Metaverse, an O-RAN interface could be a promising direction as it allows network operators to update network functions according to new applications and tasks in the Metaverse. Since machine learning has been adopted in 3GPP as a promising tool for developing network functions, improving the scalability and generalization ability of learning-based network functions remains an open issue. Note that the training of learning-based network functions may lead to huge energy consumption, we need to reconsider the energy efficiency of the whole network. Finally, privacy, security, and trust are critical for the Metaverse, where data is shared among different network functions.

\bibliography{ref}

%
%



\begin{IEEEbiographynophoto}{Zhen Meng} (z.meng.1@research.gla.ac.uk)
received his B.Eng. degree from the School of Engineering, University of Glasgow, UK, in 2019. He is currently pursuing his Ph.D. degree at the University of Glasgow, UK. His research interests include the ultra-reliable and low-latency communications, communication-robotics co-design, Metaverse, and cyber-physical systems.
\end{IEEEbiographynophoto}
\begin{IEEEbiographynophoto}{Changyang She} (shechangyang@gmail.com)
		is a lecturer-level research fellow at the University of Sydney. He is a recipient of Australian Research Council Discovery Early Career Researcher Award 2021. His research interests lie in the areas of ultra-reliable and low-latency communications, wireless artificial intelligence, and Metaverse.
\end{IEEEbiographynophoto}
	
\begin{IEEEbiographynophoto}{Guodong Zhao} (guodong.zhao@glasgow.ac.uk) is a Senior Lecturer in the James Watt School of Engineering at the University of Glasgow, UK. He is an IEEE Senior Member and the senior academic lead of the Scotland 5G Centre.  His research interests are in the areas of artificial intelligence, communications,   robotics, and Metaverse. 
\end{IEEEbiographynophoto}
\begin{IEEEbiographynophoto}{Muhammad Ali Imran} (muhammad.imran@glasgow.ac.uk) is a full professor in communication systems and the head of the Autonomous System and Connectivity (ASC) Research Division in the James Watt School of Engineering at University of Glasgow, UK. He is the founding member of the Scotland 5G Centre with expertise in 5G technologies for industrial and robotics applications.
\end{IEEEbiographynophoto}

\begin{IEEEbiographynophoto}{Mischa Dohler} (mischa.dohler@ericsson.com)
  is now Chief Architect at Ericsson Inc. in Silicon Valley, working on cutting-edge topics of 6G, Metaverse, XR, Quantum and Blockchain. He serves on the Technical Advisory Committee of the FCC and on the Spectrum Advisory Board of Ofcom. He is a Fellow of the IEEE, the Royal Academy of Engineering, the Royal Society of Arts (RSA), and the Institution of Engineering and Technology (IET). He is a Top-1\% Cited Innovator across all science fields globally.
 \end{IEEEbiographynophoto}

\begin{IEEEbiographynophoto}{Yonghui Li} (yonghui.li@sydney.edu.au)
is a Professor and Director of Wireless Engineering Laboratory at the University of Sydney. He is the recipient of the Australian Queen Elizabeth II Fellowship in 2008 and the Australian Future Fellowship in 2012. He is a Fellow of IEEE. His research interests are in the areas of millimeter wave communications, machine to machine communications, coding techniques and cooperative communications. 
\end{IEEEbiographynophoto}	

\begin{IEEEbiographynophoto}{Branka Vucetic} (branka.vucetic@sydney.edu.au)
is an ARC Laureate Fellow and Director of the Centre of IoT and Telecommunications at the University of Sydney. Her current research work is in wireless networks and IoT. She is a Life Fellow of IEEE, the Australian Academy of Technological Sciences and Engineering and the Australian Academy of Science.
 \end{IEEEbiographynophoto}

\end{document}